# Ab initio study of beryllium-decorated fullerenes for hydrogen storage

Hoonkyung Lee,[1,*] Bing Huang,[2] Wenhui Duan,[2] and Jisoon Ihm[1]

[1]Department of Physics and Astronomy, FPRD, Seoul National University, Seoul, 151-747, Korea

[2]Department of Physics, Tsinghua University, Beijing 100084, People's Republic of China

ABSTRACT

We have found that a beryllium (Be) atom on nanostructured materials with $H_2$ molecules generates a Kubas-like dihydrogen complex [H. Lee et al. arXiv:1002.2247v1 (2010)]. Here, we investigate the feasibility of Be-decorated fullerenes for hydrogen storage using ab initio calculations. We find that the aggregation of Be atoms on pristine fullerenes is energetically preferred, resulting in the dissociation of the dihydrogen. In contrast, for boron (B)-doped fullerenes, Be atoms prefer to be individually attached to B sites of the fullerenes, and a maximum of one $H_2$ molecule binds to each Be atom in a form of dihydrogen with a binding energy of ~0.3 eV. Our results show that individual dispersed Be-decorated B-doped fullerenes can serve as a room-temperature hydrogen storage medium.

# I. INTRODUCTION

Hydrogen storage in solid-state materials is of importance for the development of hydrogen fuel-cell powered vehicles and for safe mobile applications.[1] Recently, nanostructured materials adsorbing hydrogen on their surface have received much attention as a hydrogen storage medium because of the potential of good reversibility, fast kinetics, and high capacity.[2-5] However, it has been found that the storage capacity in these nanomaterials decreases significantly near room temperature and ambient pressure[6]. The reason is that the binding energy of $H_2$ molecules on these materials is mediated by the van der Waals interaction (~0.04 eV) which is much short of the required energy of ~0.3−0.4 eV.[7,8]

In recent years, transition metal (TM)-dihydrogen complexes[9] have been of intense interest for hydrogen storage purposes because of the potential of the application of these complexes to hydrogen storage materials that operate at room temperature and ambient pressure. Density functional theory (DFT) studies have shown that, using TM-dihydrogen complexes, TM (Sc, Ti, V, and Ni)-decorated nanomaterials adsorb $H_2$ molecules with a binding energy of ~0.2−0.8 eV via the hybridization of TM $d$ orbits and $H_2$ $\sigma$ or $\sigma^*$ orbits.[10-15] The hydrogen storage capacity reaches the gravimetric goal of 9 wt% of the Department of Energy (DOE) by the year 2015.[16] However, it has turned out that the aggregation of TM atoms is energetically preferred and changes significantly the adsorption nature of the $H_2$ molecules, resulting in the dissociation of $H_2$ molecules and the reduction of dihydrogen complexes.[7,17,18] Instead of TM atoms, searching for different elements for decoration is necessary to overcome the issues of the aggregation in metal-decorated hydrogen storage systems. Recently, alkali or alkali earth metal atoms have been considered as elements for decorating nanomaterials for hydrogen storage. It has been found that alkali (or earth) metal-decorated nanomaterials can adsorb $H_2$ molecules with a binding energy of ~0.1−0.2 eV.[19-23]

More recently, it has been found that, similarly to TM-dihydrogen complexes, beryllium atom (Be) with $H_2$ molecules forms Be-dihydrogen complexes on nanostructures, e.g., fullerenes or carbon nanotubes.[24] It has also been found that the binding mechanism of $H_2$ molecules on the Be atom comes from the hybridization of the Be *s* or *p* orbits with the H$_2$ σ orbits. Using the equilibrium grand partition function, it has been demonstrated that H$_2$ molecules adsorb on Be atoms on nanostructures at 25 $^{\circ}$C and 30 atm and they are desorbed at 100 $^{\circ}$C and 3 atm. The binding and releasing of H$_2$ molecules at the practical conditions are attributed to the desirable energy of H$_2$ molecules ~0.2−0.4 eV/H$_2$. It has been suggested that Be-dihydrogen complexes may be applicable to a nanostructured hydrogen storage material that operates at room temperature and ambient pressure.

In this paper, we show the feasibility of Be-decorated fullerenes as a hydrogen storage medium. We find that the aggregation of Be atoms on pristine fullerenes is energetically preferred and changes the adsorption nature of H$_2$ molecules, resulting in the dissociation of H$_2$ molecules and the reduction of the number of adsorbed H$_2$ molecules. In contrast, Be atoms prefer to be individually attached to B sites of B-doped fullerenes, and one H$_2$ molecule binds to each Be atom with a binding energy of ~0.3 eV. We also exhibit that a Be-decorated B-doped fullerene adsorbs H$_2$ molecules with the gravimetric capacity of 1.6 wt %.

## II. COMPUTATIONAL DETAILS

All our calculations were performed using the density functional theory with the plane-wave-based total energy minimization[25]. The exchange correlation energy functional of generalized gradient approximation (GGA)[26] was used, and the kinetic energy cutoff was taken to be 400 eV. The optimized atomic positions were relaxed until the Hellmann-Feynman force on each atom met less than 0.01 eV/Å. The supercell[27] calculations throughout were employed where the adjacent molecules were separated by over 10 Å to eliminate spurious interactions between periodic images on the different molecules.

## III. RESULTS AND DISCUSSION

It has been found that a Be atom prefers to be attached on top of the carbon-carbon bond of pristine fullerenes, and each Be atom binds up to two H$_2$ molecules.[24] In contrast, the Be atom prefers to be attached to the hexagonal center including B atoms of B-doped fullerenes and binds up to one H$_2$ molecule. Here, we investigate the question of whether Be atoms are aggregated on a fullerene (C$_{60}$) and a B-doped fullerene ($C_{48}B_{12}$) as well as the effects of the aggregation of Be atoms on the adsorption of $H_2$ molecules. Figures 1(a)−1(c) show the optimized geometries of the attachment for two Be atoms on a $C_{60}$ as different adsorption configurations. We confirm that one Be atom aggregated on top of each of two carbon atoms of a hexagon of $C_{60}$ (Figure 1(c)) is energetically preferred over the other cases for the attachment of two Be atoms aggregated on top of the carbon-carbon bond of one hexagon (Figure 1(b)) and two isolated Be atoms adsorbed on the carbon-carbon bond (Figure 1(a)) by 0.10 and 1.58 eV, respectively. These results show that Be atoms prefer to be aggregated on fullerenes because the binding energy of a Be atom on fullerenes (~0.7 eV) is much smaller than the cohesive energy of bulk Be (~3.3 eV). In contrast, when two Be atoms are individually dispersed on a $C_{48}B_{12}$ (Figure 1(d)), the energy is 0.87 and 1.11 eV lower, respectively, than the other cases for the attachment of one Be atom aggregated on each of two hexagonal centers and two Be atoms aggregated on top of the carbon atoms of a hexagon as shown in Figs. 1(e) and 1(f), respectively. This suppression of the aggregation of Be atoms is attributed to preferential binding of Be atoms to B sites of the $C_{48}B_{12}$ with a large binding energy of ~3 eV per Be atom.

The attractive feature is that, like the case of an isolated Be atom on $C_{60}$,[24] up to 4 $H_2$ molecules adsorb on the two aggregated Be atoms (2 $H_2$ per Be) with the binding energy of 0.25 eV/$H_2$ as shown in Fig. 1(g). The Be-dihydrogen complexes on the two aggregated Be atoms are analogous to $Be_2^{2+}$-dihydrogen complexes in vacuum, i.e., $Be_2H_4^{2+}$.[28] The number of adsorbed $H_2$ molecules per Be atom as well as the bond length of $H_2$ molecules (~0.80 Å) in two cases is the same. In addition, the distance between the Be atom and the H atom in the Be atom on a $C_{60}$ and $Be^{+2}$ ion is 1.53 and 1.62 Å, respectively. For the Be atom attached to $C_{48}B_{12}$, up to one $H_2$ molecule adsorbs on the Be atom with the binding energy of 0.31 eV/$H_2$ as shown in Fig. 2(h) where the distance between the Be atom and the H atom is 1.68 Å and the bond length of the $H_2$ molecule is slightly elongated from 0.75 Å of the isolated molecule to 0.78 Å. The difference in the number of adsorbed $H_2$ molecules between above both cases comes from the different hybridizations between Be atom and pristine fullerenes or B-doped fullerenes.[24] The calculated values with the local density approximation (LDA) are approximately twice as much as those values presented above with the GGA. We think that the correct values are expected to be somewhere in between the two approximations.

Next, we examine how four Be atoms are dispersed on a $C_{60}$ and a $C_{48}B_{12}$ to further understand the tendency for Be aggregation. We find that, similarly to the case of two Be atoms on a $C_{60}$, four aggregated Be atoms on one hexagon of $C_{60}$ as displayed in Fig. 2(c) are energetically preferred to two aggregated Be atoms on top of each of two hexagons and four isolated Be atoms on the $C_{60}$ as shown in Figs. 2(a) and 2(b) by 1.00 and 3.95 eV, respectively. In contrast, for the $C_{48}B_{12}$, individual attachment of four Be atoms on the hexagons including B atoms (two B atom per hexagon) (Figure 2(d)) is energetically more favorable than the other cases for the attachment of four aggregated Be atoms on one hexagon and two groups of one Be atom aggregated on each of two adjacent hexagonal centers by 2.07 and 2.33 eV as shown in Figs. 2(e) and 2(f), respectively. Therefore, Be atoms prefer to be individually dispersed on B-doped fullerenes as opposed to be aggregated on pristine fullerenes.

We also examine the effects of the aggregation of four Be atoms on adsorption of $H_2$ molecules. Figure 3(a) shows eight $H_2$ molecules put on the four aggregated Be atoms (2 $H_2$ per Be) with a distance of ~1.7 Å between the $H_2$ molecules and the Be before the energy minimization calculation is carried out. We find that the aggregation of Be changes the adsorption nature of $H_2$ molecules compared to the case of an isolated Be atom on a $C_{60}$, resulting in the dissociation of $H_2$ molecules and the reduction of the number of adsorbed $H_2$ molecules as displayed in Fig. 3(b). The bonding lengths and geometry among aggregated Be atoms are changed by the adsorption of $H_2$ molecules. Therefore, the number of adsorbed $H_2$ molecules as well as the binding energy of the $H_2$ molecules is reduced. The behavior is similar to aggregation effects of transition metal atoms on other nanomaterials.[7,17,18] In contrast, individually dispersed Be atoms on a $C_{48}B_{12}$ adsorb $H_2$ molecules as in the case of a single Be atom attached to the $C_{48}B_{12}$ without any changes. Therefore, B-doped fullerenes may be suitable for a backbone for Be decoration because of no aggregation of Be atoms.

Figure 3(c) exhibits the optimized geometry of a Be-decorated $C_{48}B_{12}$ with maximally adsorbed $H_2$ molecules. Since the attachment of Be atoms on the hexagons including two B atoms is preferred and the $C_{48}B_{12}$ has the six hexagons, only adsorption of six Be atoms is considered. The hydrogen storage gravimetric capacity is 1.6 wt% of hydrogen, and the molecular formula for the Be-decorated $C_{48}B_{12}$ is expressed as $C_{48}B_{12} \cdot 6Be \cdot 6H_2$. The capacity would further increase if the number of Be atoms decorating the $C_{48}B_{12}$ is increased.

## IV. CONCLUSION

In conclusion, we have performed ab initio calculations for Be dispersion on fullerenes and adsorption of $H_2$ molecules to search for a hydrogen storage medium consisting of Be atoms and fullerenes. Like transition metal-decorated nanomaterials, individually dispersed Be-decorated fullerenes can serve as a hydrogen storage medium that operates at room temperature and ambient pressure. Be atoms prefer to be individually attached on B-doped fullerenes, and $H_2$ molecules bind to each Be atom with a binding energy of ~0.3 eV/$H_2$. We feel that our results stimulate further theoretical efforts to the search for hydrogen storage media in a combination of Be atoms and various nanostructures such as carbon nanotubes and graphene.

**ACKNOWLEDGMENT.** This research was supported by the Core Competence Enhancement Program (2E21040) of KIST through the Hybrid Computational Science Lab, and the Korean Government MOEHRD, Basic Research Fund No. KRF-2006-341-C000015. The work at Tsinghua was supported by the Ministry of Science and Technology of China (Grant Nos. 2006CB605105, 2006CB0L0601), the Natural Science Foundation of China.

*E-mail: hkiee3@snu.ac.kr

REFERENCES

1. L. Schlapbach and A. Züttel, Nature (London) **414**, 353 (2001).

2. A. C. Dillon, K. M. Jones, T. A. Bekkedahl, C. H. Kiang, D. S. Bethune, and M. J. Heben, Nature (London) **386**, 377 (1997).

3. N. L. Rosi, J. Eckert, M. Eddaoudi, D. T. Vodak, J. Kim, M. O'Keeffe, and O. M. Yaghi, Science **300**, 1127 (2003).

4. S. S. Kaye and J. R. Long, J. Am. Chem. Soc. **127**, 6506 (2005).

5. S. S. Han, H. Furukawa, O. M. Yaghi, and W. A. Goddard III, J. Am. Chem. Soc. **130**, 11580 (2008).

6. Y. Ye, C. C. Ahn, C. Witham, B. Fultz, J. Lui, A. G. Rinzler, D. Colbert, K. A. Smith, and R. E. Smally, Appl. Phys. Lett. **414**, 343 (2001).

7. H. Lee, W. I. Choi, M. C. Nguyen, M.-H. Cha, E. Moon, and J. Ihm, Phys. Rev. B **76**, 195110 (2007).

8. Y.-H. Kim, Y. Zhao, A. Williamson, M. J. Heben, and S. B. Zhang, Phys. Rev. Lett. **96**, 016102 (2006).

9. G. J. Kubas, J. Organomet. Chem. **635**, 37 (2001).

10. Y. Zhao, Y.-H. Kim, A. C. Dillon, M. J. Heben, and S. B. Zhang, Phys. Rev. Lett. **94**, 155504 (2005).

11. T. Yildirim and S. Ciraci, Phys. Rev. Lett. **94**, 175501 (2005).

12. H. Lee, W. I. Choi, and J. Ihm, Phys. Rev. Lett. **97**, 056104 (2006).

13. W. H. Shin, S. H. Yang, W. A. Goddard III, and J. K. Kang, Appl. Phys. Lett. **88**, 053111 (2006).

14. N. Park, S. Hong, G. Kim, and S.-H. Jhi, J .Am. Chem. Soc. **129**, 8999 (2007).

15. G. Kim, S.-H. Jhi, and N. Park, Appl. Phys. Lett. **92**, 013106 (2008).

16. http://www.eere.energy.gov/hydrogenandfuelcells/mypp.

17. Q. Sun, Q. Wang, P. Jena, and Y. Kawazoe, J. Am. Chem. Soc. **127**, 14582 (2005).

18. S. Li and P. Jena, Phys. Rev. Lett. **97**, 209601 (2006).

19. H. Lee, B. Huang, W. Duan, and J. Ihm, Appl. Phys. Lett. **93**, 063107 (2008).

20. Q. Sun, Q. Wang, and P. Jena, Appl. Phys. Lett. **94**, 013111 (2009).

21. C. Ataca, E. Aktürk, S. Ciraci, and H. Ustunel, Appl. Phys. Lett. **93**, 043123 (2008).

22. H. Lee, J. Ihm, M. L. Cohen, and S. G. Louie, Nano Lett. **10**, 793 (2010).

23. Y. Y. Sun, K. Lee, Y.-H. Kim, and S. B. Zhang, Appl. Phys. Lett. **95**, 033109 (2009).

24. H. Lee, B. Huang, W. Duan, and J. Ihm, arXiv:1002.2247v1 (2010).

25. J. Ihm, A. Zunger, and M. L. Cohen, J. Phys. C **12**, 4409 (1979).

26. J. P. Perdew and Y. Wang, Phys. Rev. B **45**, 13244 (1992).

27. M. L. Cohen, M. Schlüter, J. R. Chelikowsky, and S. G. Louie, Phys. Rev. B **12**, 5575 (1975).

28. C. A. Nicolaides and P. Valtazanos, Chem. Phys. Lett. **172**, 254 (1990).

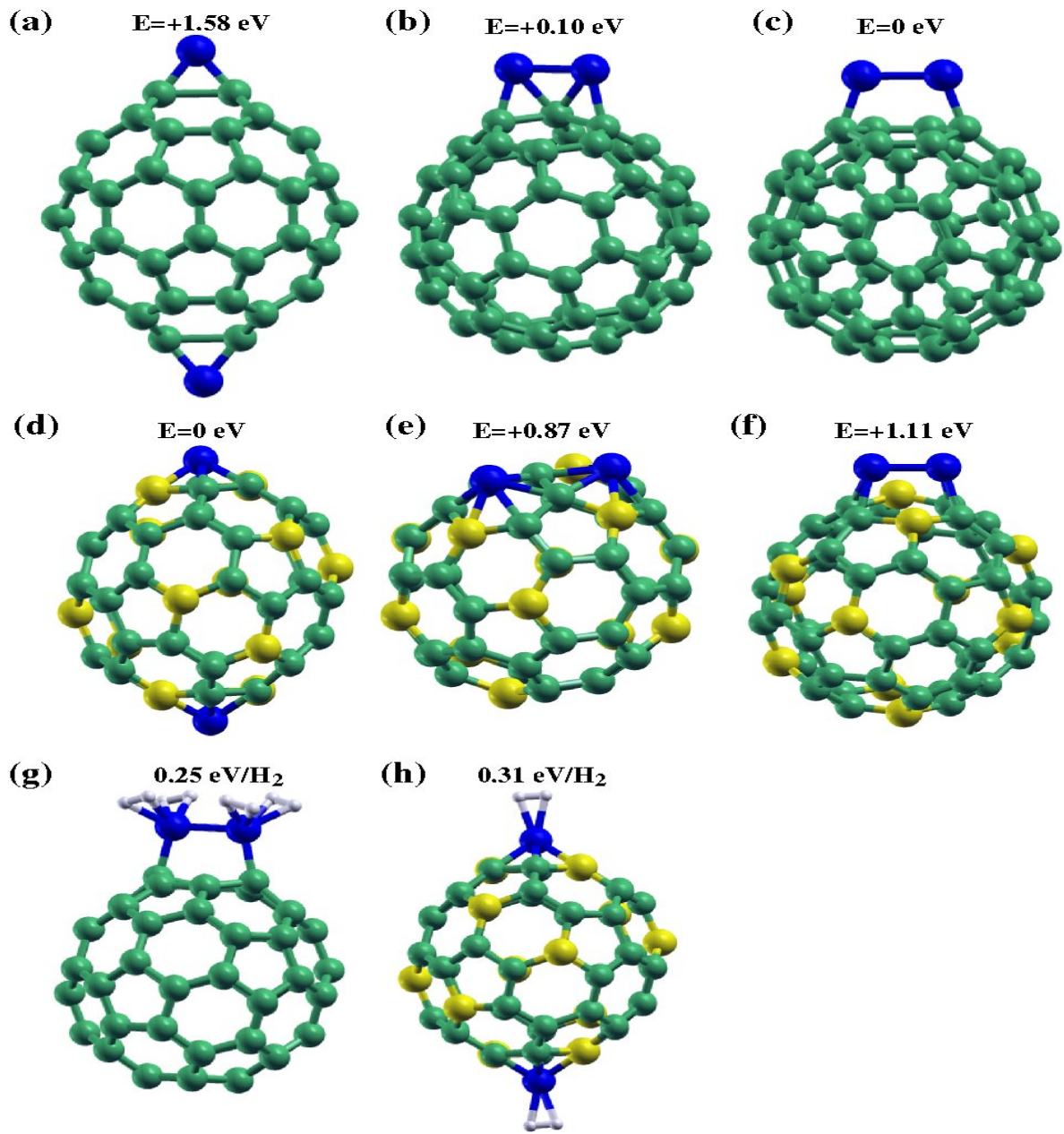

**FIG. 1**. (a) Two Be atoms individually attached on top of two of the carbon-carbon bonds of a $C_{60}$. (b) Two Be atoms aggregated on top of the carbon-carbon bond of a hexagon of a $C_{60}$. (c) Two Be atoms aggregated on top of the carbon atoms of a hexagon of a $C_{60}$. (d) One Be atom individually attached on each of two hexagonal centers including two B atoms of a $C_{48}B_{12}$. (e) One Be atom aggregated on each of two hexagonal centers of a $C_{48}B_{12}$. (f) Two Be atoms aggregated on top of the carbon atoms of a hexagon of a $C_{48}B_{12}$. (g) and (h) Optimized geometries for the adsorption of four and two $H_2$ molecules on two aggregated and one isolated Be atoms on $C_{60}$, respectively. The total energy (E) of the lowest energy structure is set to zero. Green, blue, yellow, and white dots indicate carbon atom, beryllium atom, boron atom, and hydrogen atom.

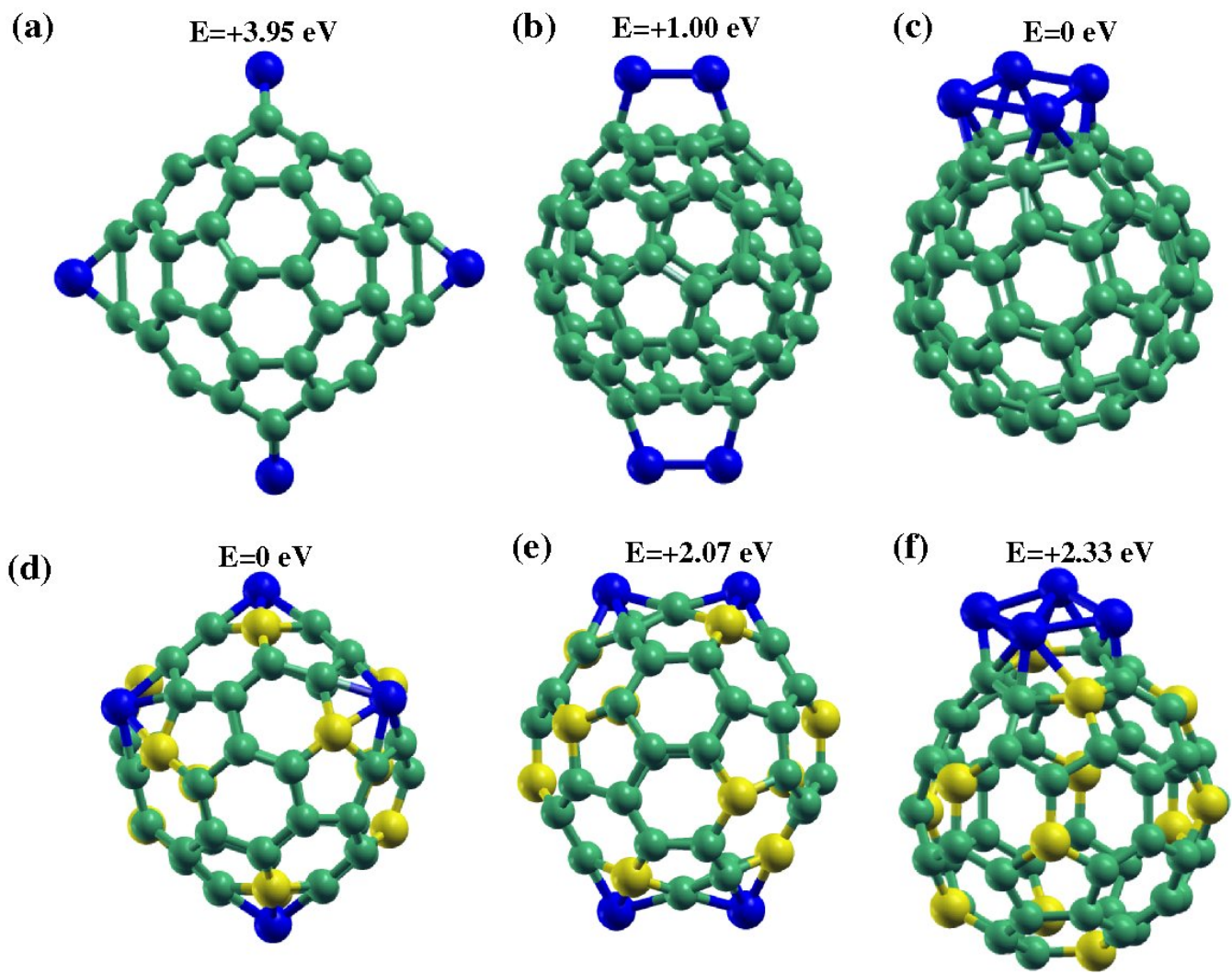

**FIG. 2**. (a) One Be atom individually attached on top of each of four carbon-carbon bonds of a $C_{60}$. (b) Two aggregated Be atoms on top of each of two hexagons of a $C_{60}$. (c) Four Be atoms aggregated on one hexagon of a $C_{60}$. (d) One Be atom individually attached on each of four hexagonal centers including two B atoms of a $C_{48}B_{12}$. (e) Two groups of one Be atom aggregated on each of two adjacent hexagonal centers of a $C_{48}B_{12}$. (f) Four Be atoms aggregated on one hexagon of a $C_{48}B_{12}$. The total energy (E) of the lowest energy structure is set to zero.

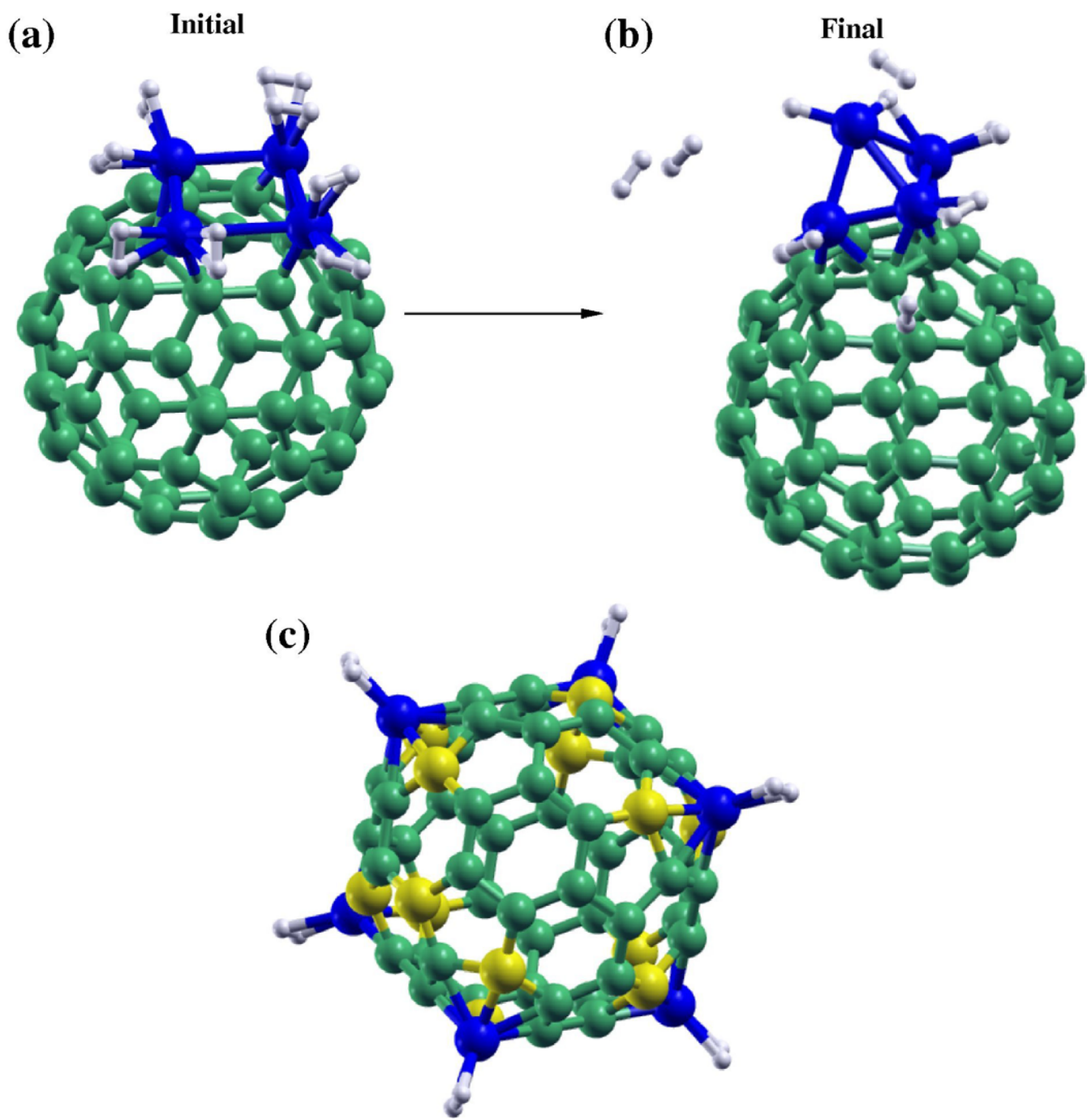

**FIG. 3**. (a) and (b) Atomic structures for the adsorption of eight $H_2$ molecules on four aggregated Be atoms (2 $H_2$ per Be) on a $C_{60}$ before and after the energy minimization calculation, respectively. (c) The optimized atomic structure for a Be-decorated $C_{48}B_{12}$ with maximally adsorbed $H_2$ molecules.